\documentclass[12pt]{article}
\usepackage{epsf}
\usepackage{amsmath}
\usepackage{amssymb}
\usepackage{graphics}
\usepackage{cite}

\renewcommand{\thefootnote}{\#\arabic{footnote}}

\begin{document}

\setcounter{footnote}{0}
\begin{titlepage}

\begin{center}

\hfill July 2008\\

\vskip .5in

{\Large \bf
Probing  the Effective Number of Neutrino Species with Cosmic Microwave Background 
}

\vskip .45in

{\large
Kazuhide Ichikawa$^{1,2}$,
Toyokazu Sekiguchi$^1$,
and Tomo Takahashi$^3$ 
}

\vskip .45in

{\em
$^1$
Institute for Cosmic Ray Research, 
University of Tokyo, Kashiwa 277-8582, Japan\\
$^2$
Department of Physics and Astronomy, University College London, Gower Street, London, WC1E 6BT, U.K. \\
$^3$
Department of Physics, Saga University, Saga 840-8502, Japan  
}

\end{center}

\vskip .4in

\begin{abstract}
  We discuss how much we can probe the effective number of neutrino
  species $N_\nu$ with cosmic microwave background alone.  
  Using the data of WMAP, ACBAR, CBI and
  BOOMERANG experiments, we obtain a constraint on the effective
  number of neutrino species as 
  $0.96< N_\nu <7.94$
  at 95\,\%\,C.L.  for a power-law $\Lambda$CDM flat universe model. 
  The limit is  improved to be 
  $1.39 < N_\nu < 6.38$
  at 95\,\%\,C.L.   if we assume that the
  baryon density, $N_\nu$ and the helium abundance are related by the
  big bang nucleosynthesis theory.  We also provide a forecast for the
  PLANCK experiment using a Markov chain Monte Carlo approach.  In
  addition to constraining $N_\nu$, we investigate how the big bang
  nucleosynthesis relation affects the estimation for these parameters
  and other cosmological parameters.
 \end{abstract}
\end{titlepage}

\renewcommand{\thepage}{\arabic{page}}
\setcounter{page}{1}
\renewcommand{\thefootnote}{\#\arabic{footnote}}

\section{Introduction}

Cosmology is now becoming a precision science, and cosmological
observations can give us a lot of information for our understanding of
the universe.  Moreover, the interplay between cosmology and particle
physics in various contexts has also been discussed vigorously.  One
of such examples is the effective number of neutrino species $N_\nu$.
Although collider experiments such as LEP have measured the number of
light active neutrino types to be $2.92 \pm 0.06$ \cite{Yao:2006px},
it is important to cross-check this value because
cosmological measurements may lead to different value.
This could be due to an extra radiation component which is predicted 
by some models of particle physics such as sterile neutrinos (see Ref.~\cite{Dolgov:2002wy} 
and references therein), or due to incomplete thermalization of neutrinos
in the low-scale reheating universe in which the reheating temperature $T_{\rm reh}$ 
can be as low as $T_{\rm reh} \sim \mathcal{O}(1)$\,MeV and $N_\nu$ is predicted to be less than three
  \cite{Kawasaki:1999na,Kawasaki:2000en,Hannestad:2004px,Ichikawa:2005vw}.
 If such a non-standard ingredient exists, it can affect big bang
nucleosynthesis (BBN), cosmic microwave background (CMB), large scale
structure (LSS) and so on; thus precise cosmological observations can
probe these scenarios through the effective number of neutrino
species.

Constraints on $N_\nu$ have been investigated in the literature using
the information of CMB and LSS, sometimes with priors on the Hubble
constant, cosmic age and Helium abundance 
\cite{Kneller:2001cd,Hannestad:2001hn,Bowen:2001in,Crotty:2003th,Pierpaoli:2003kw,Hannestad:2003xv,Barger:2003zg,Crotty:2004gm,Hannestad:2005jj,Spergel:2006hy,Seljak:2006bg,Hannestad:2006mi,Cirelli:2006kt,Ichikawa:2006vm,Mangano:2006ur,Friedland:2007vv,Hamann:2007pi,Ichikawa:2007fa,deBernardis:2007bu,Popa:2008nz,Komatsu:2008hk,Dunkley:2008ie,Simha:2008zj,Popa:2008tb}.
Although CMB in general can constrain various quantities severely,
since the effects of $N_\nu$ on CMB are degenerate with some
cosmological parameters, the studies so far have combined CMB data
with some other observations such as LSS to obtain a sensible
constraint on $N_\nu$.  However, when one uses the data from LSS,
constraints can become different depending on how one treats
non-linear correction/bias on small scales for the matter power
spectrum \cite{Hamann:2007pi}.  Furthermore, different LSS data seem
to give different constraints on $N_\nu$
\cite{Spergel:2006hy,Seljak:2006bg,Ichikawa:2006vm,Mangano:2006ur,Hamann:2007pi}.
Regarding the prior on the Hubble constant $H_0$, as is summarized in
Ref.~\cite{Ichikawa:2006vm}, it can yield some constraints on $N_\nu$
when combined with CMB data (without LSS data)
\cite{Hannestad:2001hn,Bowen:2001in,Crotty:2003th,Hannestad:2003xv,Barger:2003zg},
but they depend on the $H_0$ prior adopted.  One may consider that we
can use the usually assumed prior on the Hubble constant based on the
result by Freedman et al. $H_0 = 72 \pm 8$ 
\cite{Freedman:2000cf}, but another group reported a
somewhat lower value as $H_0 = 62.3 \pm 5.2 $ \cite{Sandage:2006cv}.  If the lower value for
$H_0$ is adopted as the prior, a resulting constraint on $N_\nu$ would
be different.  Having these considerations in mind, it is desirable to
investigate a constraint on $N_\nu$ without these kind of
uncertainties.

In this paper, we study a constraint on $N_\nu$ from CMB experiments
alone.  By making the analysis of CMB data alone, we can avoid such
subtleties as the galaxy-bias/non-linear corrections and the value for
the prior on the Hubble constant.  However, as is mentioned above, the
effects of $N_\nu$ are strongly degenerate in CMB with other
cosmological parameters such as energy density of matter, the Hubble
constant, and the scalar spectral index, and, in fact, we could not obtain
a meaningful bound only with WMAP3
\cite{Ichikawa:2006vm,Hamann:2007pi}.
Recent WMAP5 alone analysis gives a better constraint but it still cannot 
give an upper bound \cite{Dunkley:2008ie,Komatsu:2008hk}.
As we will discuss later, the
degeneracy is significant up to about the 2nd/3rd peak of the CMB
power spectrum where the observation of WMAP has precisely measured.
To break this degeneracy to some extent, it would be helpful to have
the information at higher multipoles where signals unique to
relativistic neutrinos are expected to appear \cite{Bashinsky:2003tk}.
Recently, the data from ACBAR
which probes CMB at higher multipoles than those of WMAP has been
updated \cite{Reichardt:2008ay}.  By using this data in addition to
other small scale observations such as BOOMERANG and CBI, we can
obtain a relatively severe constraint on $N_\nu$ which is comparable
to that have been obtained previously with LSS data.

The organization of this paper is as follows. In the next section, we
start with the discussion how $N_\nu$ affects the CMB power spectrum,
which helps to understand our results for the constraint on $N_\nu$. In
Section \ref{sec:current}, we study the current constraint on $N_\nu$
using observations of CMB alone. We use the data from WMAP5, the
recent ACBAR, BOOMERANG and CBI. Furthermore, we forecast the
constraint from the future Planck experiment.  In the final section,
we summarize our results and discuss its implications for some models
of particle physics/the early universe.

\section{Effects of $N_\nu$ on CMB}\label{sec:effects}

The effective number of neutrino species $N_\nu$ represents the energy
density stored in relativistic components as
\begin{equation}
\label{eq:N_nu}
\rho_{\rm rad} = \rho_\gamma + \rho_\nu  + \rho_x 
= \left[ 1 + \frac{7}{8} \left(\frac{4}{11}\right)^{4/3} N_\nu \right] \rho_\gamma
\end{equation}
where $\rho_\gamma$, $\rho_\nu$ and $\rho_x$ are energy densities of
photons, three species of massless active neutrinos and some possible
extra radiation components, respectively. In this paper, we assume
that neutrinos are massless and have no chemical potential.  For the
case with the standard three neutrino flavors without an extra
relativistic component, the effective number of neutrino is $N_\nu =
3.046$ where some corrections from the incomplete decoupling due to a
slight interaction of neutrinos with electrons/positrons and finite
temperature QED effect to the electromagnetic plasma are taken into
account \cite{Mangano:2005cc}.  Any deviation of $N_\nu$ from this
value implies that there exists an extra relativistic component and/or
some non-standard thermal history takes place such as the low
reheating temperature scenario.

\begin{figure}[htb]
\begin{center}
\scalebox{1}{
\includegraphics{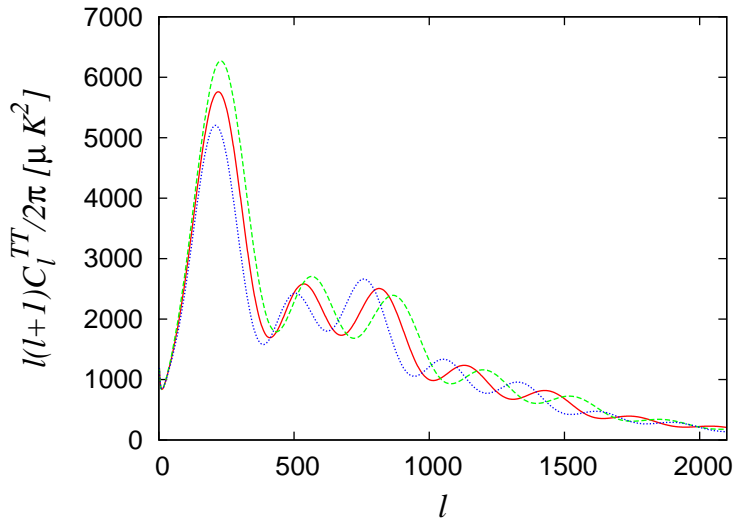}
}
\caption{CMB power spectra for the cases with $N_\nu=1$ (blue dotted
  line), $3$ (red solid line) and $5$ (green dashed line). Other
  cosmological parameters are taken as the mean value from WMAP5 alone
  analysis for a power-law flat $\Lambda$CDM model.  }
\label{fig:cl}
\end{center}
\end{figure}

To illustrate the effects of $N_\nu$ on CMB, we plot CMB power spectra
for several values of $N_\nu$ in Fig.~\ref{fig:cl}.  Other
cosmological parameters are assumed as the mean values of WMAP5 alone
analysis for a power-law flat $\Lambda$CDM model.  As seen from the
figure, as we increase the value of $N_\nu$, the height of the 1st
peak is strongly enhanced and the positions of acoustic peaks are
shifted to higher multipoles.  Also, the amplitude on small scales
(higher multipoles) is suppressed.  Below, we discuss where these
changes are coming from.

One of the main effects of $N_\nu$ comes from the change of the epoch
of the radiation-matter equality.  By increasing (decreasing) the
value of $N_\nu$, the radiation-matter equality occurs later (earlier).
 Thus the increase (decrease) of $N_\nu$ gives an almost the
same effect of the decrease (increase) of energy density of matter.
One of noticeable features is that the height of the first acoustic
peak is enhanced by increasing the value of $N_\nu$.  This is due to
the early integrated Sachs-Wolfe (ISW) effect in which fluctuations of
the corresponding scale, having crossed the sound horizon in the
radiation-dominated epoch are boosted by the decay of the gravitational
potential.  Thus a larger amount of relativistic species drives the
first peak higher.  Another effect is the shift of the position of
acoustic peaks due to the change of the radiation-matter equality
through the change of $N_\nu$. The position of acoustic peaks is well
captured by the so-called acoustic scale $\theta_A$ which is inversely
proportional to the peak position and written as
\begin{equation}
\label{eq:theta_A}
\theta_A = \frac{r_s (z_{\rm rec})}{r_\theta (z_{\rm rec})}
\end{equation}
where $r_\theta (z_{\rm rec})$ and $r_s (z_{\rm rec})$ are the
comoving angular diameter distance to the last scattering surface and
the sound horizon at the recombination epoch $z_{\rm rec}$,
respectively. Although $r_\theta (z_{\rm rec})$ almost remains the
same for different values of $N_\nu$, $r_s (z_{\rm rec})$ becomes
smaller when $N_\nu$ is increased.  Thus the positions of acoustic
peaks are shifted to higher multipoles (smaller scales) by increasing
the value of $N_\nu$. Furthermore, since the position of the $n$-th
peak can be roughly written as $l_n \sim n \pi / \theta_A$,
separations of the peaks become also greater for larger $N_\nu$.

Another important effect is free-streaming of ultrarelativistic
neutrinos \cite{Bashinsky:2003tk}.  The perturbation of
ultrarelativistic neutrino propagates with the speed of light, which
is faster than the sound speed of acoustic oscillations of
photon-baryon fluid.  The coupled photon-baryon component behaves to
oscillate like a compressional fluid; on the other hand,
ultrarelativistic neutrinos free-stream to erase their fluctuations.
These two components are coupled via gravity; thus the fluctuations of
photons can also be affected by the free-streaming of neutrinos, which
results in the damping of the amplitude and the shift of the acoustic
oscillations.  The effects are significant on small scales where
fluctuations of a given scale enter the horizon while the energy density
of ultrarelativistic neutrinos takes a significant portion of that of
the universe.

Although the effects of the standard cosmological parameters on the
heights and positions of acoustic peaks are well known, here we
discuss them in some phenomenological way including the effects of
$N_\nu$.  For this purpose, we calculated the response of the heights
and positions of the acoustic peaks to the change of the cosmological
parameters up to the 5th peak around the fiducial values. As a
fiducial parameter set, we take those of the mean value from WMAP5-alone 
analysis for a power-law flat $\Lambda$CDM model.  
The shifts of the positions of acoustic peaks are
\begin{eqnarray}
\Delta l_1 &=& 
15.58\frac{\Delta \omega_b}{\omega_b} 
-26.99\frac{\Delta \omega_m}{\omega_m} 
+36.01\frac{\Delta n_s}{n_s} 
+0.94\frac{\Delta Y_p}{Y_p}
-44.58\frac{\Delta h}{h} 
+15.53\frac{\Delta N_\nu}{N_\nu} ,
\label{eq:l1} \\
\Delta l_2 &=& 
62.57\frac{\Delta \omega_b}{\omega_b} 
-74.90\frac{\Delta \omega_m}{\omega_m} 
+14.69\frac{\Delta n_s}{n_s} 
+2.81\frac{\Delta Y_p}{Y_p}
-108.60\frac{\Delta h}{h} 
+47.73\frac{\Delta N_\nu}{N_\nu} ,
\label{eq:l2} \\
\Delta l_3 &=& 
74.23\frac{\Delta \omega_b}{\omega_b} 
-143.47\frac{\Delta \omega_m}{\omega_m} 
+9.88\frac{\Delta n_s}{n_s} 
+4.64\frac{\Delta Y_p}{Y_p}
-152.39\frac{\Delta h}{h} 
+81.82\frac{\Delta N_\nu}{N_\nu} ,
\label{eq:l3} \\
\Delta l_4 &=& 
110.84\frac{\Delta \omega_b}{\omega_b} 
-181.89\frac{\Delta \omega_m}{\omega_m} 
+7.29\frac{\Delta n_s}{n_s} 
+6.77\frac{\Delta Y_p}{Y_p}
-220.21\frac{\Delta h}{h} 
+112.76\frac{\Delta N_\nu}{N_\nu} ,
\label{eq:l4} \\
\Delta l_5 &=& 
136.88\frac{\Delta \omega_b}{\omega_b} 
-237.18\frac{\Delta \omega_m}{\omega_m} 
+6.20\frac{\Delta n_s}{n_s} 
+7.79\frac{\Delta Y_p}{Y_p}
-276.02\frac{\Delta h}{h} 
+145.34\frac{\Delta N_\nu}{N_\nu} ,
\label{eq:l5} 
\end{eqnarray}
where $l_i$ means the position of the $i$-th peak.  $\omega_b$,
$\omega_m$, $n_s$, $Y_p$ and $h$ are energy densities of baryon and
matter, the scalar spectral index, the primordial helium abundance and
the normalized Hubble constant.  
$n_s$  is defined at the wave number $k = 0.05$\,Mpc$^{-1}$.
The positive derivatives of the peak
positions with respect to $N_\nu$ demonstrate the decrease in $r_s
(z_{\rm rec})$ due to the increase in $N_\nu$.

The responses of the heights of acoustic peaks to the change of various
cosmological parameters are
\begin{eqnarray}
\frac{\Delta \mathcal{C}_{l_1}}{\mathcal{C}_{l_1}} &=& 
0.429\frac{\Delta \omega_b}{\omega_b} 
-0.632\frac{\Delta \omega_m}{\omega_m} 
-0.947\frac{\Delta n_s}{n_s} 
-0.0065\frac{\Delta Y_p}{Y_p}
+0.141\frac{\Delta N_\nu}{N_\nu}, 
\label{eq:Cl1} \\
\frac{\Delta  \mathcal{C}_{l_2}}{\mathcal{C}_{l_2}} &=& 
-0.211\frac{\Delta \omega_b}{\omega_b} 
-0.579\frac{\Delta \omega_m}{\omega_m} 
-0.034\frac{\Delta n_s}{n_s} 
-0.035\frac{\Delta Y_p}{Y_p}
+0.083\frac{\Delta N_\nu}{N_\nu} ,
\label{eq:Cl2} \\
\frac{\Delta  \mathcal{C}_{l_3}}{\mathcal{C}_{l_3}} &=& 
0.026\frac{\Delta \omega_b}{\omega_b} 
-0.136\frac{\Delta \omega_m}{\omega_m} 
+0.276\frac{\Delta n_s}{n_s} 
-0.071\frac{\Delta Y_p}{Y_p}
-0.080\frac{\Delta N_\nu}{N_\nu} ,
\label{eq:Cl3} \\
\frac{\Delta  \mathcal{C}_{l_4}}{\mathcal{C}_{l_4}} &=& 
-0.044\frac{\Delta \omega_b}{\omega_b} 
-0.229\frac{\Delta \omega_m}{\omega_m} 
+0.587\frac{\Delta n_s}{n_s} 
-0.125\frac{\Delta Y_p}{Y_p}
-0.108\frac{\Delta N_\nu}{N_\nu} ,
\label{eq:Cl4} \\
\frac{\Delta  \mathcal{C}_{l_5}}{\mathcal{C}_{l_5}} &=& 
0.149\frac{\Delta \omega_b}{\omega_b} 
-0.006\frac{\Delta \omega_m}{\omega_m} 
+0.776\frac{\Delta n_s}{n_s} 
-0.172\frac{\Delta Y_p}{Y_p}
-0.216\frac{\Delta N_\nu}{N_\nu}. \label{eq:Cl5_ratio} 
\end{eqnarray}
where ${\cal C}_l = l(l+1)C^{TT}_l/2\pi$.
Since $h$ gives only a negligible change to the height of the peaks, we omit it.
By increasing the value of $N_\nu$, the height of the 1st peak is
enhanced due to the early ISW effect, as previously mentioned.  For
the third and the higher peaks, the heights are damped more by increasing
$N_\nu$, which is inferred from the negative coefficients.  This
is due to the effect of free streaming of neutrinos
\cite{Bashinsky:2003tk}.

In addition, for later convenience, we also show the derivatives of the
peak heights relative to the first peak height following
Refs.~\cite{Hu:2000ti}. Here, $H_i \equiv {\cal C}_i/{\cal C}_1$ for
$i=2$--5.  They are useful quantities when we interpret degeneracies
since the dependence on the overall amplitude is cancelled out.
\begin{eqnarray}
\Delta H_2 &=& 
-0.291\frac{\Delta \omega_b}{\omega_b} 
+0.023\frac{\Delta \omega_m}{\omega_m} 
+0.396\frac{\Delta n_s}{n_s} 
-0.013 \frac{\Delta Y_p}{Y_p}
-0.026\frac{\Delta N_\nu}{N_\nu} ,
\label{eq:H2} \\
\Delta H_3 &=&
-0.177\frac{\Delta \omega_b}{\omega_b} 
+0.206\frac{\Delta \omega_m}{\omega_m} 
+0.514\frac{\Delta n_s}{n_s} 
-0.028 \frac{\Delta Y_p}{Y_p}
-0.098\frac{\Delta N_\nu}{N_\nu} , 
\label{eq:H3} \\
\Delta H_4 &=& 
-0.102\frac{\Delta \omega_b}{\omega_b} 
+0.082\frac{\Delta \omega_m}{\omega_m} 
+0.317\frac{\Delta n_s}{n_s} 
-0.025 \frac{\Delta Y_p}{Y_p}
-0.054\frac{\Delta N_\nu}{N_\nu} , 
\label{eq:H4} \\
\Delta H_5 &=& 
-0.040\frac{\Delta \omega_b}{\omega_b} 
+0.084\frac{\Delta \omega_m}{\omega_m} 
+0.236\frac{\Delta n_s}{n_s} 
-0.023 \frac{\Delta Y_p}{Y_p}
-0.052\frac{\Delta N_\nu}{N_\nu}.
\label{eq:H5} 
\end{eqnarray}

\section{Constraint on $N_\nu$ from observations of CMB}\label{sec:current}
In this section, we present our result for the constraint 
on $N _\nu$ from CMB alone.  First, we give some details of
our analysis. We use the CMB data of 
WMAP5 \cite{Komatsu:2008hk,Dunkley:2008ie,Hinshaw:2008kr,Hill:2008hx,Nolta:2008ih},
BOOMERANG \cite{Jones:2005yb,Piacentini:2005yq,Montroy:2005yx}, CBI
\cite{Sievers:2005gj} and ACBAR \cite{Reichardt:2008ay}.  
We performed a Markov chain Monte
Calro (MCMC) analysis to obtain constraints on cosmological parameters
using {\tt cosmomc} code \cite{Lewis:2002ah} with some modifications
which are described in the following.  We explore a 9 dimensional
parameter space which consists of $\omega_b$, $\omega_c$, $\tau$,
$\theta_s$, $n_s$, $A_s$, $A_{SZ}$, $Y_p$ and $N_\nu$. Here, $\omega_c$ is the
energy density of dark matter, $\tau$ is the optical depth of
reionization, $\theta_s$ is the acoustic peak scale
\cite{Kosowsky:2002zt},  $A_s$ is the amplitude of primordial
fluctuations 
and $A_{SZ}$ is the amplitude of thermal Sunyaev-Zel'dovich (SZ) effect 
which is normalized to the $C_l^{SZ}$ template from Ref.~\cite{Komatsu:2002wc}.

\begin{figure}[htb]
\begin{center}
\scalebox{0.9}{
\includegraphics{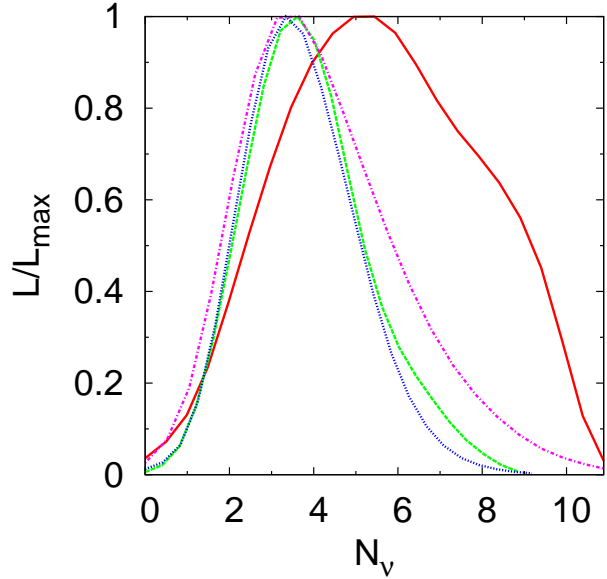}
}
\caption{1D posterior distributions of $N_\nu$. The red solid line
  uses WMAP5 alone (with $Y_p=0.24$ fixed) and the other lines use
  WMAP5+ACBAR+BOOMERANG+CBI with different assumptions on $Y_p$.  
   The  green dashed line fixes it to be $Y_p=0.24$, the blue dotted line
  uses the BBN relation to fix $Y_p$ from $\omega_b$ and $N_\nu$, and
  the magenta dot-dashed line treats $Y_p$ as a free parameter.  
  For the analysis with WMAP5 alone, we assumed the prior on the cosmic age 
  as  $10~{\rm Gyr} < t_0 < 20~{\rm Gyr}$.}
\label{fig:1Dcurrent}
\end{center}
\end{figure}

\begin{table}[ht]
\begin{center}
  \begin{tabular}{l||r|r|r}
  \hline
  \hline
  & \raisebox{-.7em}[0pt][0pt]{Mean} & 68\%$\uparrow$ & 95\%$\uparrow$ \\
  & & 68\%$\downarrow$ & 95\%$\downarrow$ \\
  \hline
  \hline
  WMAP5  & \raisebox{-.7em}[0pt][0pt]{5.65} & $7.88$ & $9.96$ \\
  ($Y_p=0.24$: fixed)& & $3.02$ & $1.92$\\
  \hline
  CMB all & \raisebox{-.7em}[0pt][0pt]{4.24} & $5.47$ & $7.94$\\
  ($Y_p$: free)& & $2.03$ & $0.96$ \\
  \hline
  CMB all & \raisebox{-.7em}[0pt][0pt]{3.71} & $4.80$ & $6.38$\\
  ($Y_p$: BBN relation)& & $2.27$ & $1.39$ \\
  \hline
  CMB all & \raisebox{-.7em}[0pt][0pt]{3.89} & $4.89$ & $6.84$\\
  ($Y_p=0.24$: fixed)& & $2.19$ & $1.28$ \\
  \hline 
  \hline 
\end{tabular}
\caption{The mean values, 68\% and 95\% limits 
    of $N_\nu$ for several current
  CMB data sets and assumptions of $Y_p$.
  }
  \label{table:N_nu_current}
\end{center}
\end{table}

\begin{figure}[htb]
\begin{center}
\begin{tabular}{cccc}
\hspace{-1.3cm}
\scalebox{0.38}{\includegraphics{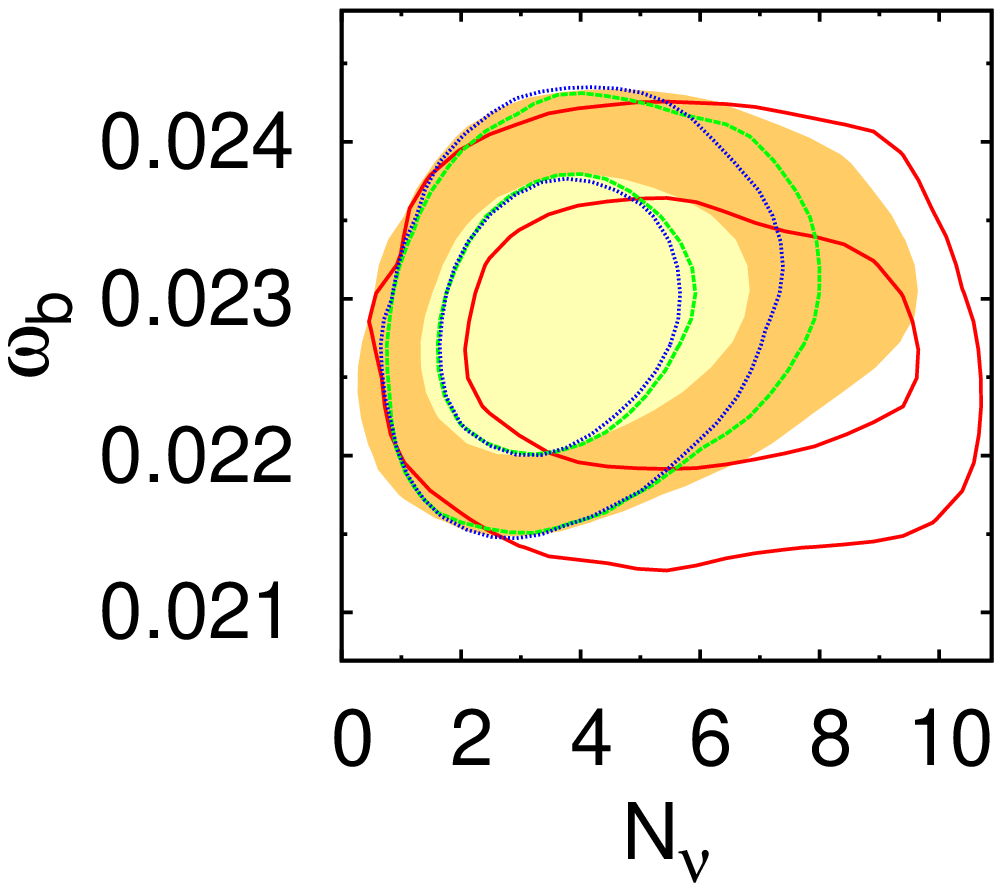}} &
\hspace{-2.1cm}
\scalebox{0.38}{\includegraphics{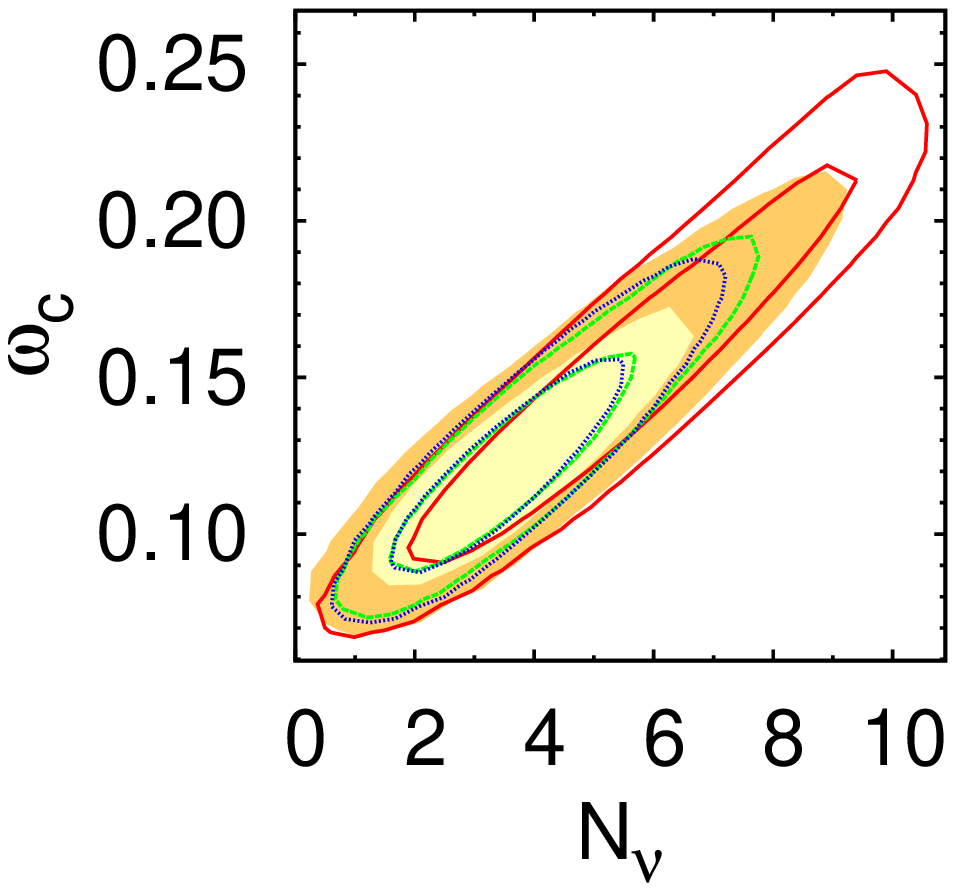}} &
\hspace{-2.1cm}
\scalebox{0.38}{\includegraphics{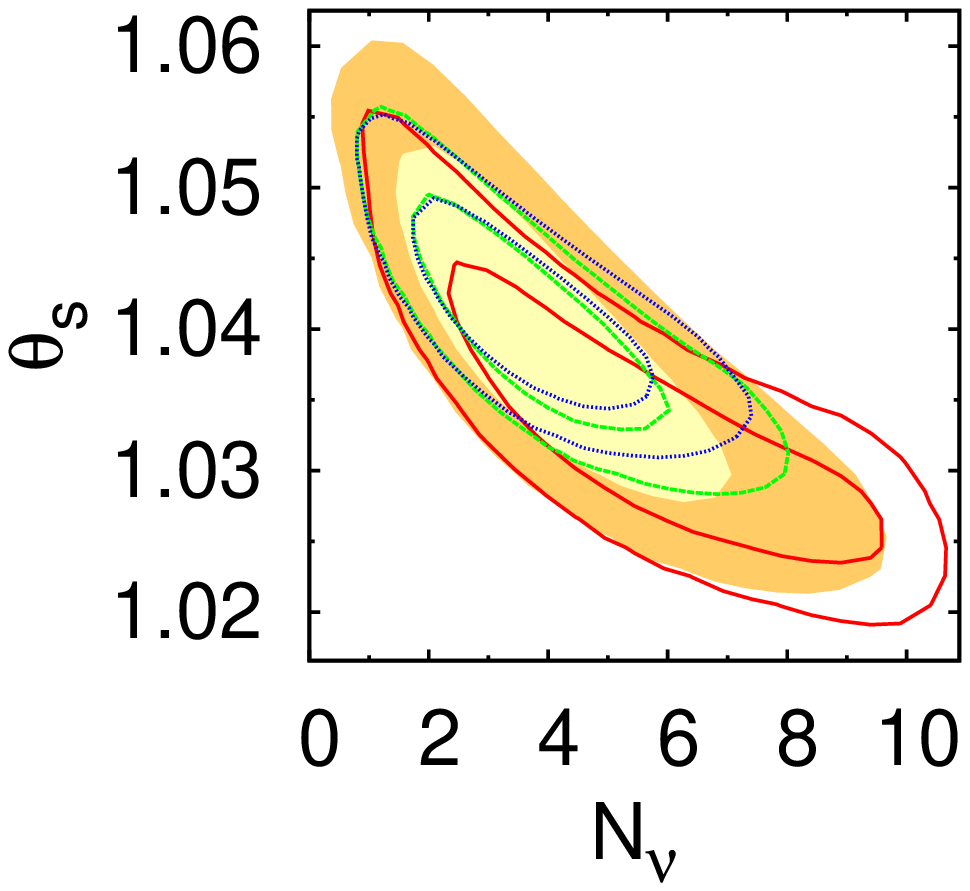}} &
\hspace{-2.1cm}
\scalebox{0.38}{\includegraphics{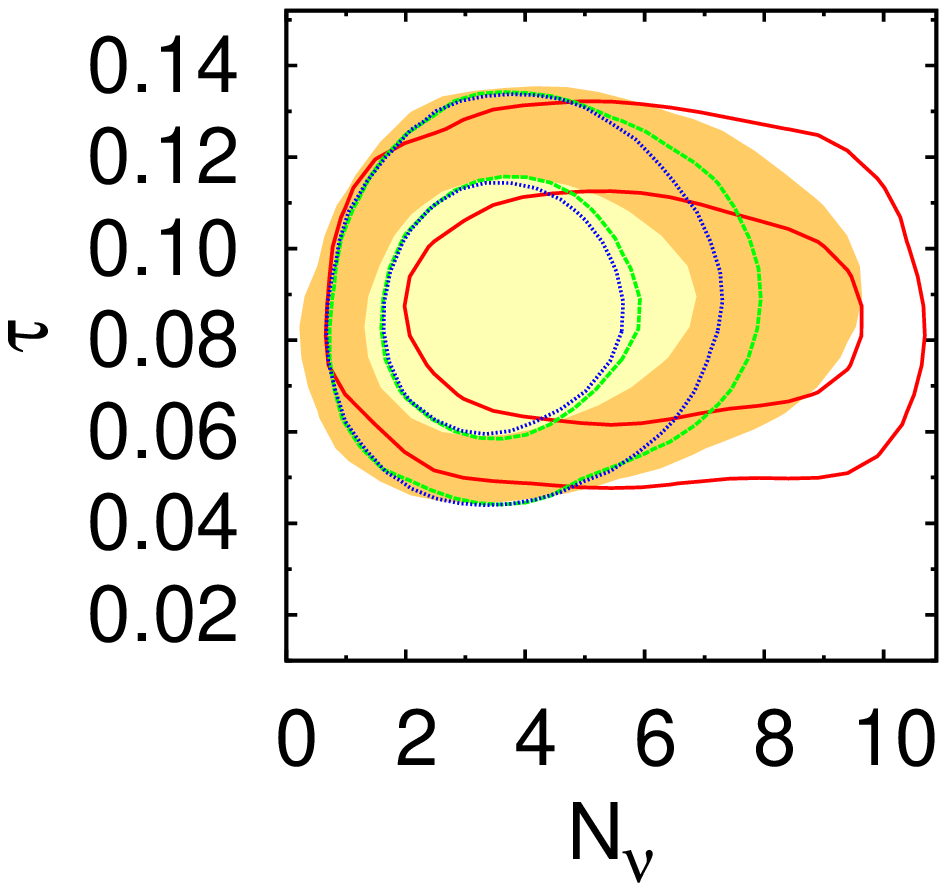}} \\
\hspace{-1.2cm}
\scalebox{0.38}{\includegraphics{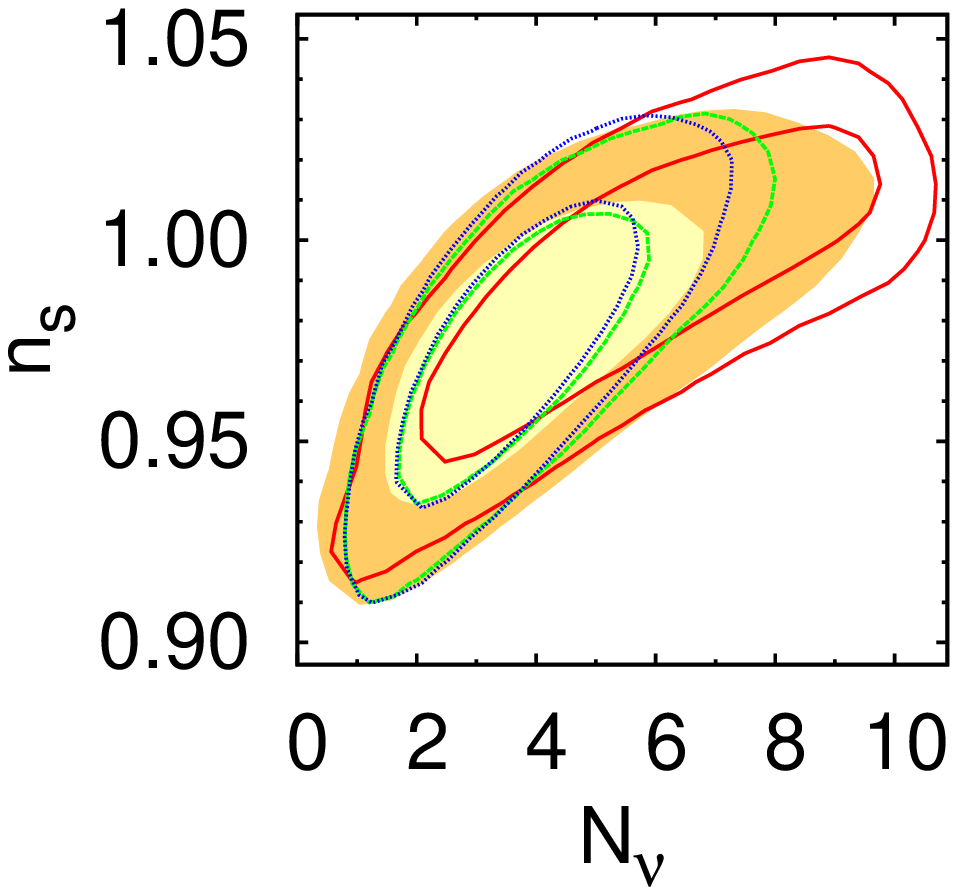}} &
\hspace{-1.9cm}
\scalebox{0.38}{\includegraphics{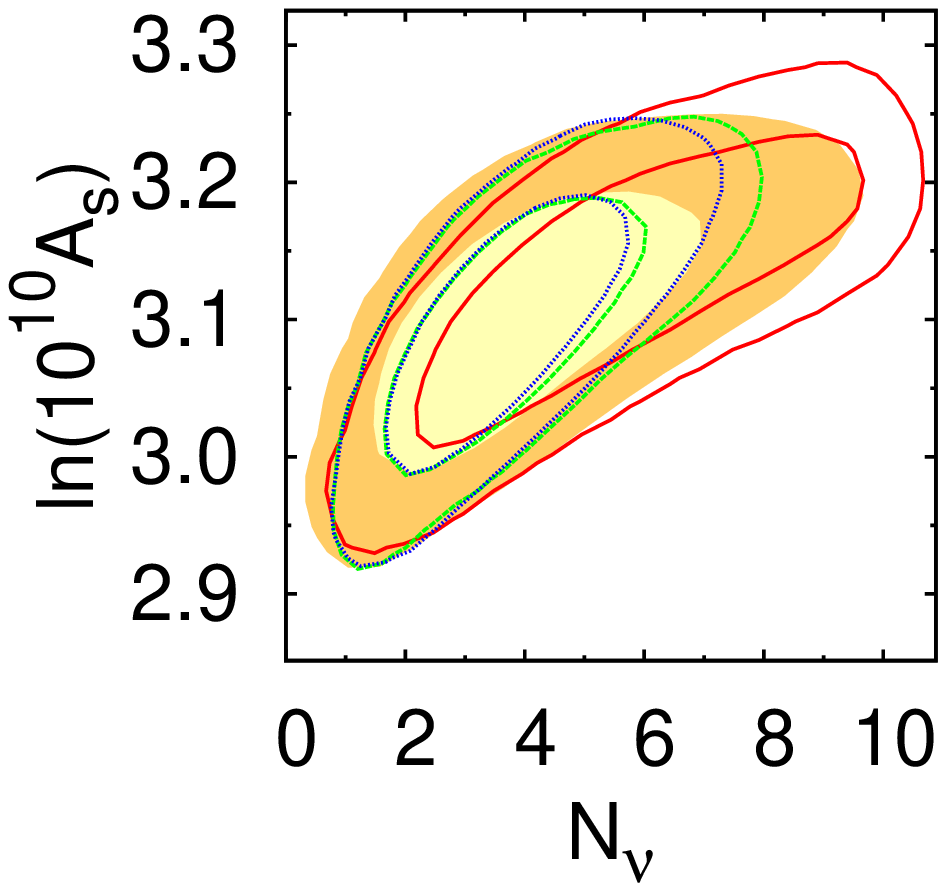}} &
\hspace{-1.9cm}
\scalebox{0.38}{\includegraphics{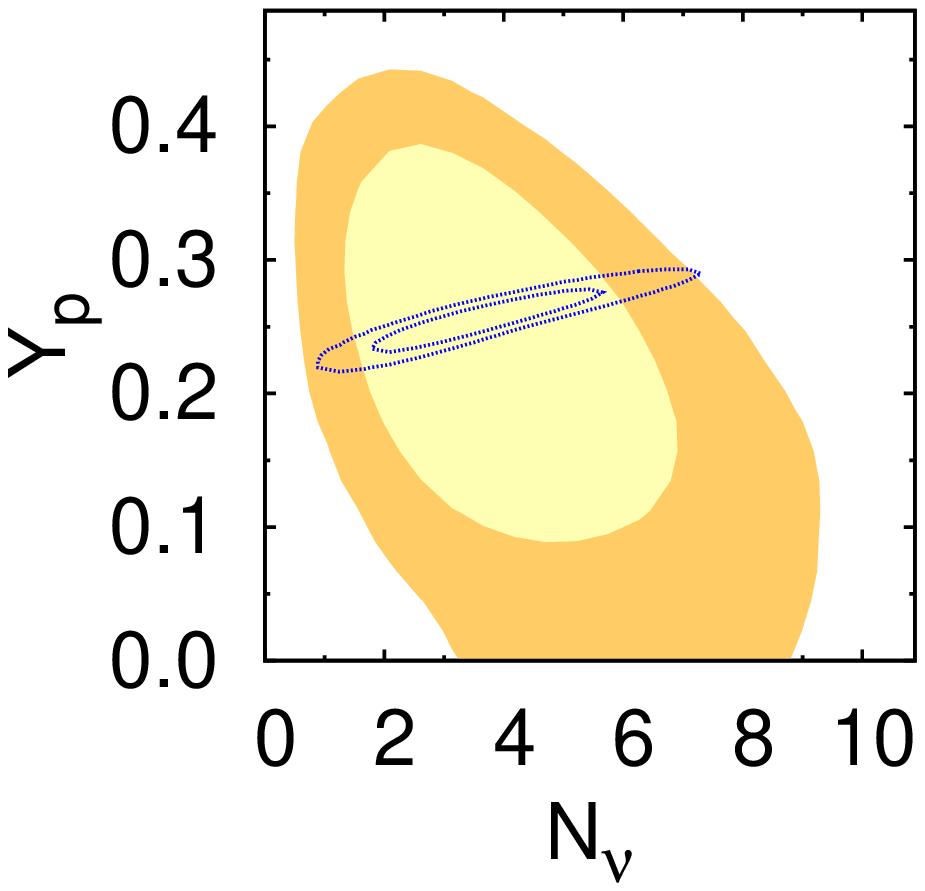}} &
\hspace{-2.1cm}
\scalebox{0.38}{\includegraphics{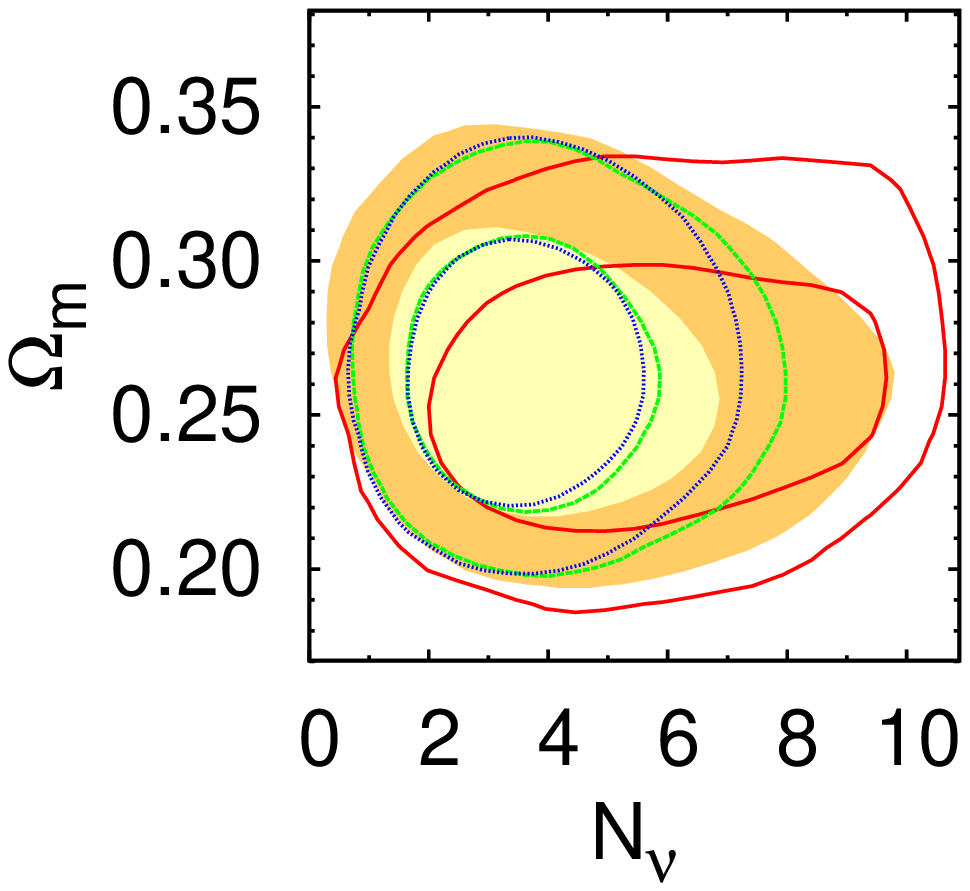}} \\
\hspace{-0.9cm}
\scalebox{0.38}{\includegraphics{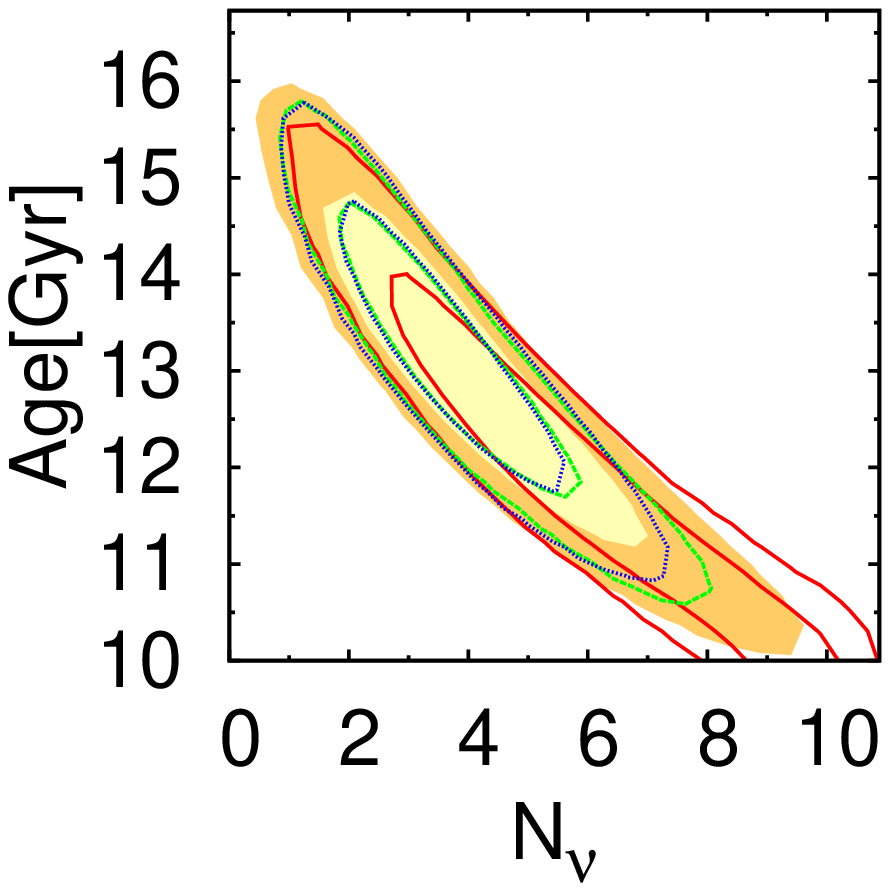}} &
\hspace{-1.9cm}
\scalebox{0.38}{\includegraphics{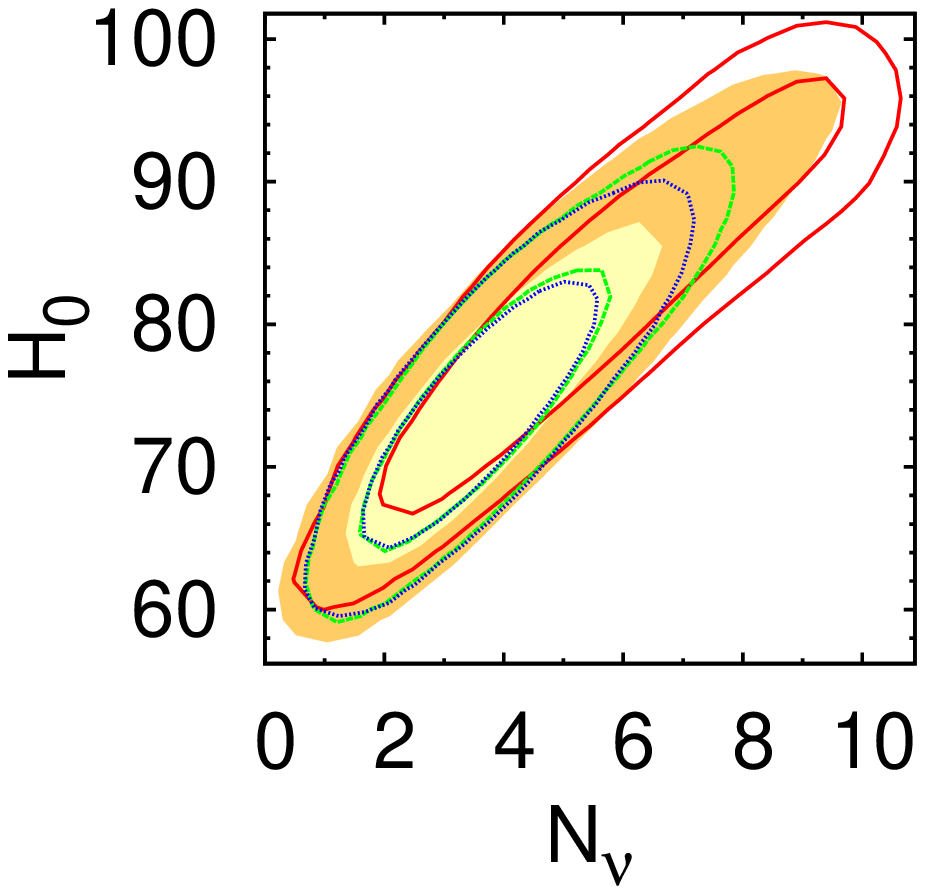}} & & \\
\end{tabular}
\caption{The 68\% and 95\% allowed regions in the plane of
  $N_\nu$ v.s.~other parameters when WMAP5 alone 
    is used with $Y_p=0.24$ (red solid line), 
   WMAP5+ACBAR+BOOMERANG+CBI are used with 
   $Y_p=0.24$ (green dashed line),  $Y_p$ being fixed by the BBN relation
   (blue dotted line) and $Y_p$ being treated as a free parameter (orange and 
   yellow shaded region).}
\label{fig:2Dcurrent}
\end{center}
\end{figure}

As far as CMB is concerned, these cosmological parameters can be
considered to be independent. However, when we take into account the
BBN theory, $Y_p$ is determined once $\omega_b$ and $N_\nu$ are given.
In this case, we should relate these parameters to each other and
sample an 8 dimensional parameter space.  We shall refer to this
relation among $Y_p$, $\omega_b$ and $N_\nu$ as the BBN relation.  For
this purpose, we calculate $Y_p$ as a function of $\omega_b$ and
$N_\nu$ using the Kawano BBN code \cite{Kawanocode} with some update
in the nuclear reaction network part based on Ref.~\cite{NACRE}. Such
relation is considered in the CMB analyses in
Refs.~\cite{Huey:2003ef,Ichikawa:2006dt,Hamann:2007sb,Ichikawa:2007js,Popa:2008nz}.
In passing, we would like to make a comment on the fitting formula for
the BBN calculation presented in Ref.~\cite{Serpico:2004gx} which has
been used in the authors' previous work \cite{Ichikawa:2006dt}. 
We do not adopt the formula here since, as we will see later, our
  MCMC chains for the constraints from the present data sets sometimes
  go to the region beyond the range over which 
their fitting formula is valid, $0 \le N_\nu \le 6$.
   For the Planck forecast, where
the chains are contained in that region, we obtain the same results
with the Kawano code and the fitting formula of
Ref.~\cite{Serpico:2004gx}.  Furthermore, we also consider the case
with fixing the helium abundance to $Y_p=0.24$ since, in most
analyses, the primordial helium abundance is fixed to this value.
Finally, it should be noted that this BBN relation is not
  necessarily realized in some cases.  We can think of more exotic
scenarios in  which the BBN theory cannot relate
  those parameters. 
For exmaple,  $\omega_b$ and $Y_p$ may vary between BBN and CMB
epochs \cite{Ichikawa:2004pb} or an increase in $N_\nu$ may take place
\cite{Ichikawa:2007jv}. 

Now, we present our results in order. In Fig.~\ref{fig:1Dcurrent}, the
posterior 1D distributions for $N_\nu$ are shown for the analysis with
WMAP5 alone and WMAP5+ACBAR+BOOMERANG+CBI (CMB all). 
The former is shown by a red solid line.
 For the latter case, the results for different assumptions on $Y_p$ are
  depicted: $Y_p$ being fixed as $Y_p=0.24$ (green dashed line), $Y_p$
  determined from the BBN relation (blue dotted line) and $Y_p$ being
  treated as a free parameter (magenta dot-dashed line).
Corresponding constraints on $N_\nu$ 
are summarized in Table \ref{table:N_nu_current}.
We also show 2D contours of 68\% and 95\% C.L. in the
planes of $N_\nu$ v.s.~several other cosmological parameters in
Fig.~\ref{fig:2Dcurrent}.  Table \ref{table:current_all} summarizes
the derived constraints on these parameters.

As seen from Fig.~\ref{fig:1Dcurrent}, the likelihood for $N_\nu$ from WMAP alone 
has irregular shape, far from Gaussian. It has the maximum
around $N_\nu \sim 5$, declines slowly as $N_\nu$ increases and drop
abruptly at $N_\nu \sim 9$.  The abrupt cut can be traced to the
  prior on the cosmic age $t_0$ which is implicitly assumed as $ 10\,
  {\rm Gyr} < t_0 < 20\, {\rm Gyr}$ in the analysis.  In particular,
the lower limit $t_0 > 10\,{\rm Gyr}$ makes the cut (see
Fig~\ref{fig:2Dcurrent}).  We can regard this prior to be very
conservative on the observational ground since it is far looser than
the astrophysical lower bound of the cosmic age e.g.~$t_0 > 11.2\,{\rm
  Gyr}$ (95\% C.L.) from the age estimates of globular clusters
\cite{Krauss:2003em}.  Moreover, we should include such prior from the
practical reason. As can be seen by the relatively slow decline of the
likelihood for $5 \lesssim N_\nu \lesssim 9$ or the elongated contours
in Fig~\ref{fig:2Dcurrent}, the degeneracy is so severe that we cannot
produce MCMC chains which are well converged within a reasonable time.
Although we can formally calculate a constraint using this data as
shown in Table \ref{table:N_nu_current}, since the likelihood is so
irregular, we would conclude that it is not meaningful to constrain $N_\nu$
from WMAP5 alone.  

However, it may be instructive to understand how the degeneracies
arise in the WMAP-alone analysis.  As clearly shown in
Fig.~\ref{fig:2Dcurrent}, $N_\nu$ most notably degenerates with
$\omega_c$ and $H_0$ (or $\theta_s$). There are also some degeneracies with $n_s$ and
$A_s$ but not as severe as $\omega_c$ or $H_0$. These degeneracies are
understood as follows. First, to produce the same amount of the early
ISW effect, $\omega_c$ has to be increased as $N_\nu$ increases. It
roughly scales as $(\omega_b+\omega_c) \propto N_\nu$ to make the
matter-radiation equality same. At the same time, under the flatness
assumption, $\Omega_m$ has to be preserved in order to have the same
distance to the last scattering surface. Then, since $\Omega_m =
(\omega_b+\omega_c)/h^2$, $h$ has to increase for larger $N_\nu$. The
slight enhancement in $n_s$ and $A_s$ can be attributed to their
effects to cancel the suppression around the diffusion damping scales
due to the increase in $N_\nu$. A more quantitative argument based on
the derivatives presented in the previous section may be useful. The
degeneracy as regards the same matter-radiation equality is given by
setting $\Delta {\cal C}_{l_1} = 0$ to be $\Delta \omega_m/\omega_m \sim
0.2\, \Delta N_\nu/N_\nu$.  This is equivalent to $\Delta
\omega_c/\omega_c \sim 0.3\, \Delta N_\nu/N_\nu$, which roughly gives
the slope in the $N_\nu$--$\omega_c$ plane in Fig~\ref{fig:2Dcurrent}.
Using this relation with $\Delta l_1=0$ shows the $N_\nu$--$h$
degeneracy. From Eq.~\eqref{eq:l1}, we obtain $\Delta h/h \sim 0.2\,
\Delta N_\nu/N_\nu$, which appear in the $N_\nu$--$h$ contour in
Fig~\ref{fig:2Dcurrent}. The $N_\nu$--$n_s$ degeneracy is given by
further requiring $\Delta H_2 = 0$. Plugging $\Delta
\omega_m/\omega_m$ and $\Delta h/h$ in Eq.~\eqref{eq:H2} yields
$\Delta n_s/n_s \sim 0.05\,\Delta N_\nu/N_\nu$. This 5\% increase in the
best fit value of $n_s$ for $\Delta N_\nu = 3$ is consistent with the
$N_\nu$--$n_s$ contour in Fig~\ref{fig:2Dcurrent}. Although WMAP has
measured the CMB power spectrum very precisely, since it is just up to
around the 2nd peak, the effects of $N_\nu$ are absorbed in the
changes of $\omega_c$, $h$, $n_s$ and $A_s$ and we cannot constrain
$N_\nu$.

\begin{figure}[htb]
\begin{center}
\scalebox{1}{
\includegraphics{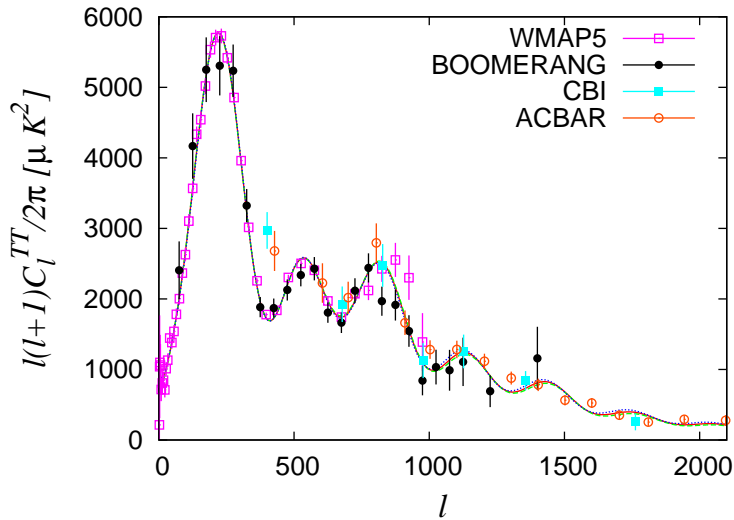}
}
\caption{An illustration of the degeneracy of $N_\nu$ with other
  cosmological parameters. Here the value of the effective number of
  neutrino are assumed as $N_\nu=1$ (blue dotted line), $3$ (red solid
  line) and $5$ (green dashed line) and other cosmological parameters
  are chosen such that CMB spectra becomes the same as that with the
  fiducial parameters. }
\label{fig:cl_deg}
\end{center}
\end{figure}

For a visual illustration of the degeneracy, in Fig.~\ref{fig:cl_deg},
we show CMB power spectra for the case with $N_\nu=1$ (blue dotted
line), $3$ (red solid line) and $5$ (green dashed line) with other
cosmological parameters being chosen such that they give the
degenerate spectra up to the 2nd/3rd peak.  We can see that these
curves coincide up to the 2nd peak but begin to separate around the 3rd or
the higher peaks.  On small scales, the change caused by $N_\nu$ cannot be
fully canceled just by tuning other parameters. In particular, $n_s$
affects the spectrum in the whole scales; thus, even if the spectra is
almost degenerate up to the 2nd peak by tuning the value of $n_s$, it
cannot cancel the damping on small scales.  

\begin{table}[ht]
  \begin{center}
  \begin{tabular}{l||r|r|r|r}
  \hline
  \hline 
  parameters & CMB all & CMB all & CMB all & WMAP5 \\
  & ($Y_p$: free) & ($Y_p$: BBN relation) & ($Y_p=0.24$) & ($Y_p=0.24$)\\
  \hline
  $\omega_b$ & $0.0229^{+0.00056}_{-0.00054}$ & $0.02291^{+0.00052}_{-0.00059}$ & 
  $0.02291^{+0.00058}_{-0.00053}$ & $0.02275^{+0.00060}_{-0.00062}$ \\
  $\omega_c$ & $0.132^{+0.018}_{-0.037}$ & $0.124^{+0.017}_{-0.027}$ & 
  $0.127^{+0.017}_{-0.029}$ & $0.153^{+0.036}_{-0.047}$\\
  $\theta_s$ & $1.0390^{+0.0071}_{-0.0088}$ & $1.0413^{+0.0039}_{-0.0051}$ & 
  $1.0402^{+0.0044}_{-0.0060}$ & $1.0334^{+0.0042}_{-0.0084}$\\
  $\tau$ & $0.088^{+0.016}_{-0.019}$ & $0.088^{+0.016}_{-0.017}$ & 
  $0.088^{+0.015}_{-0.019}$ & $0.088^{+0.016}_{-0.019}$\\
  $n_s$ & $0.977^{+0.026}_{-0.022}$ & $0.975^{+0.024}_{-0.023}$  & 
  $0.975^{+0.024}_{-0.021}$ & $0.989^{+0.030}_{-0.019}$\\
  $\ln(10^{10}A_s)$ & $3.104^{+0.0067}_{-0.0061}$ & $3.097^{+0.066}_{-0.061}$ & 
  $3.098^{+0.065}_{-0.059}$ & $3.128^{+0.080}_{-0.054}$\\
  $Y_p$ & $0.220^{+0.105}_{-0.085}$ & $0.256^{+0.015}_{-0.016}$ & 
  --- & ---\\
  $N_\nu$ & $4.24^{+1.23}_{-2.21}$ & $3.72^{+1.07}_{-1.45}$ & 
  $3.89^{+1.00}_{-1.70}$ & $5.65^{+2.63}_{-2.23}$ \\
  $A_\mathrm{SZ}$ & $1.04^{+0.96}_{-0.34}$ & $1.07^{+0.93}_{-0.33}$ & 
  $1.05^{+0.95}_{-0.34}$ & $1.00^{+0.85}_{-0.50}$\\
  \hline
  $\Omega_m$ & $0.265^{+0.026}_{-0.032}$ & $0.266^{+0.025}_{-0.030}$ & 
  $0.265^{+0.025}_{-0.030}$ & $0.260^{+0.028}_{-0.033}$\\
  Age[Gyr] & $12.9^{+1.3}_{-1.2}$ & $13.2^{+1.0}_{-1.0}$ & 
  $13.1^{+1.0}_{-1.1}$ & $12.1^{+7.7}_{-1.9}$\\
  $H_0$ &  $76.5^{+6.4}_{-9.7}$ & $74.5^{+5.5}_{-6.7}$ & 
  $75.2^{+5.6}_{-7.3}$ & $82.0^{+9,9}_{-8.9}$\\
  \hline
  \hline 
\end{tabular}
  \caption{Mean values and 68\% C.L. errors from current observations of CMB
  for the cases with WMAP5 alone and all data combined.}
  \label{table:current_all}
\end{center}
\end{table}

When we include the data at higher multipoles such as ACBAR, BOOMERANG
and CBI, the likelihood functions now have a well-behaved peak close to
Gaussian and we can obtain meaningful constraints.  The bound on $N_\nu$ is 
$0.96\le N_\nu \le 7.94$ 
at 95 \% C.L.~ when $Y_p$ is treated as a
free parameter.  At higher multipoles, the free streaming of neutrinos
damps the spectrum, which cannot be compensated by above-mentioned
parameters.  Hence the degeneracy can be removed to some extent. 
That is the reason why we can have a severer constraint on $N_\nu$ by including the
data on small scales.  In fact, $Y_p$ also suppresses the amplitude on
small scales via diffusion damping (see Eqs.~\eqref{eq:Cl4} and \eqref{eq:Cl5_ratio}); 
thus the constraint on $N_\nu$
slightly changes for different treatments of $Y_p$ but the differences
are very small as seen in Fig.~\ref{fig:1Dcurrent} and
Table~\ref{table:N_nu_current}.  Since current CMB
observations on small scales are not so precise, it does not make much
difference how we treat $Y_p$. Imposing the BBN relation tightens the
constraint to 
$1.39 \le N_\nu \le 6.38$ 
at 95 \% C.L., but it is not
so different from the $Y_p$-free case. Also, the limit does not
differ much even if we set $Y_p = 0.24$.  Similarly, the estimates for
the other cosmological parameters are not affected by the assumption on 
$Y_p$ as shown in Table~\ref{table:current_all}.

\begin{table}[ht]
\begin{center}
  \begin{tabular}{l||r|r|r||r|r||r|r}
  \hline
  \hline
  & \multicolumn{3}{c||}{No priors on $N_\nu$} & \multicolumn{2}{c||}{$N_\nu \ge 3.046$} & \multicolumn{2}{c}{$N_\nu \le 3.046$} \\
  \hline
  & \raisebox{-.7em}[0pt][0pt]{Mean} & 68\%$\uparrow$  & 95\%$\uparrow$ & 68\%$\uparrow$ & 95\%$\uparrow$ & 68\%$\uparrow$ & 95\%$\uparrow$ \\
  & & 68\%$\downarrow$ & 95\%$\downarrow$ & 68\%$\downarrow$ & 95\%$\downarrow$ & 68\%$\downarrow$ & 95\%$\downarrow$\\ 
  \hline
  \hline
  CMB all & \raisebox{-.7em}[0pt][0pt]{$4.24$} & $5.47$ & $7.94$ & $5.51$ & $8.19$ & (3.046) & (3.046) \\
  ($Y_p$: free) & & $2.03$ & $0.96$ & $(3.046)$ & $(3.046)$ & $2.05$ & $1.17$\\
  \hline
  CMB all & \raisebox{-.7em}[0pt][0pt]{$3.71$} & $4.80$ & $6.38$ & $4.70$ & $6.35$ & \multicolumn{2}{r}{\raisebox{-.7em}[0pt][0pt]{---}}\\
  ($Y_p$: BBN relation) & & $2.27$ & $1.39$ & $(3.046)$ & $(3.046)$ & \multicolumn{2}{l}{}\\
  \hline
  CMB all & \raisebox{-.7em}[0pt][0pt]{$3.89$} & $4.89$ & $6.84$ & $4.87$ & $6.88$ & (3.046) & (3.046) \\
  ($Y_p=0.24$: fixed) & & $2.19$ & $1.28$ & $(3.046)$ & $(3.046)$ & $2.12$ & $1.27$\\
  \hline
  \hline
  Planck & \raisebox{-.7em}[0pt][0pt]{$3.11$} & $3.44$ & $3.83$ & $3.45$ & $3.87$ & (3.046) & (3.046) \\
  ($Y_p$: free) & & $2.72$ & $2.41$ & $(3.046)$ & $(3.046)$ & $2.72$ & $2.43$\\
  \hline 
  Planck & \raisebox{-.7em}[0pt][0pt]{$3.06$} & $3.26$ & $3.44$ & $3.25$ & $3.44$ & \multicolumn{2}{r}{\raisebox{-.7em}[0pt][0pt]{---}}\\
  ($Y_p$: BBN relation) & & $2.87$ & $2.68$ & $(3.046)$ & $(3.046)$ & \multicolumn{2}{l}{}\\
  \hline
  Planck & \raisebox{-.7em}[0pt][0pt]{$3.19$} & $3.43$ & $3.67$ & $3.37$ & $3.63$ & (3.046) & (3.046) \\
  ($Y_p=0.24$: fixed) & & $2.95$ & $2.72$ & $(3.046)$ & $(3.046)$ & $2.87$ & $2.67$\\
  \hline
  \hline 
\end{tabular}
  \caption{The mean values and 68\% and 95\% limits of $N_\nu$ for current and future
  CMB data.
  }
  \label{table:N_nu}
\end{center}
\end{table}

Up to now, we have assumed no prior on $N_\nu$.  However, if we
consider an extra radiation component such as sterile neutrinos and so
on, the effective number of neutrino species just increases.  In this
case, $N_\nu$ cannot be less than the standard value of 3.046.  Thus
it may be appropriate to study adopting the prior $N_\nu > 3.046$ to
constrain a scenario with such an extra radiation component.
We denote it as $\Delta N^{\rm ext}_\nu \equiv N_\nu  - 3.046$.
With this prior, we obtain an upper bound on an extra radiation component as 
$N_\nu <8.19$ (or $\Delta N^{\rm ext}_\nu < 5.14$) 
at 95 \% C.L. when $Y_p$ is taken as a
free parameter and improve to be 
$N_\nu < 6.35$ (or $\Delta N^{\rm ext}_\nu < 3.30$) 
at 95 \% C.L. when the BBN relation is assumed.
Notice that these limits are weaker than those with no priors on $N_\nu$,
which are
$N_\nu < 7.94$ ($\Delta N^{\rm ext}_\nu < 4.89$) and 
$N_\nu < 6.38$ ($\Delta N^{\rm ext}_\nu < 3.33$), respectively.
This somewhat peculiar fact stems from the shape of the likelihood 
shown in Fig.~\ref{fig:1Dcurrent},
which is not symmetric with respect to $N_\nu = 3.046$. 
Since the differences due to the $N_\nu$ prior are not negligible, 
caution is needed when we use these constraints 
regarding the prior on $N_\nu$.

Even when we limit ourselves to the case with three active neutrino
species, a deviation from the standard value of $N_\nu = 3.046$ is
possible.  In a scenario with low (MeV scale) reheating temperature,
$N_\nu$ can be less than 3.046. 
  In this case, $N_\nu$ only takes
the value less than the standard one.  Thus it may be interesting to
investigate a constraint on $N_\nu$ assuming $N_\nu < 3.046$.
As regards the treatment of  $Y_p$, we do not consider the case with
adopting the BBN relation here because, in a scenario with MeV
reheating temperature, $Y_p$ should be calculated taking into account
the non-thermal neutrino distribution functions and oscillation
effects
\cite{Kawasaki:1999na,Kawasaki:2000en,Hannestad:2004px,Ichikawa:2005vw}.
These effects drive $Y_p$ to increase as $N_\nu$ decreases contrary to
the usual case where $N_\nu$ just represents a measure of the expansion
rate. 
(This is why we are not showing constraints for the prior $N_\nu < 3.046$ with
the BBN relation in Table \ref{table:N_nu}.)
Since taking into account this effect is beyond the scope of this paper,
we show the constraint for the case with $Y_p$ being varied freely,
which can be considered as the conservative one for the prior $N_\nu < 3.046$.
We obtained constraints $N_\nu > 1.27$ and $N_\nu > 1.17$ at 95\% C.L. for the cases with
$Y_p$ being fixed as $Y_p=0.24$ and $Y_p$ being assumed as a free
parameter, respectively. For a scenario with MeV scale reheating temperature,
these limits are translated into the lower bound on the reheating
temperature as $T_{\rm reh}>2.0$\,MeV \cite{Ichikawa:2005vw}. 

Finally, we investigate a future constraint on cosmological parameters
paying particular attention to $N_\nu$ and its effects on constraints
on other parameters. We use the data expected from the future Planck
experiment and make a MCMC analysis following the method in Ref.~\cite{Perotto:2006rj}.  
As for the specification of
Planck, we adopt the following parameters for the instrument. For the
frequency channels of $\nu = 100$, 143 and 217\,GHz, the width of the
beam and the sensitivities per pixel for temperature and polarization
are adopted as $(\theta_{\rm FWHW} {\rm [arcmin]}, \sigma_T [\mu {\rm
  K}], \sigma_P [\mu {\rm K}] ) =(9.5, 6.8, 10.9), (7.1, 6.0, 11.4)$
and $(5.0, 13.1, 26.7)$, respectively.  Other frequency channels are
assumed to be used to remove foregrounds.  
We make use of the data up to
$l =2500$ in order that our results will not be affected by the SZ effect 
and the marginalization over $A_{SZ}$ is not performed.
These setups for the Planck forecast are similar to the recent works
performed in Refs.~\cite{Hamann:2007sb,Popa:2008nz,Popa:2008tb},  but explored 
parameter spaces are different. We make a simple extension by adding
$N_\nu$ and $Y_p$ to the standard 6 dimensional parameter 
space, but theirs include neutrino masses and/or
lepton asymmetry.  When one would like to check the constraint 
on an extra radiation component in a simple scenario,
one can refer our results here. However, when some other particular 
setups are considered such as a scenario with large lepton asymmetry 
and massive neutrinos,
the above mentioned references should be consulted.

Our results are summarized in Tables~\ref{table:N_nu} and 
\ref{table:forecast}. 
As seen from the Table~\ref{table:N_nu}, the constraint is most 
stringent when the BBN relation is adopted,
 and in this case,  we obtained a future constraint as 
 $2.68 \le N_\nu \le3.44$ 
 at 95 \% C.L.  Another point which should be noted is that 
fixing of $Y_p = 0.24$ can bias the determination of 
some other cosmological parameters 
such as $\omega_b$ and $n_s$, which was already pointed out in 
Refs.~\cite{Hamann:2007sb,Ichikawa:2007js}.
However, when we vary the value of $N_\nu$, 
the effect of fixing of $Y_p=0.24$ is partly cancelled by the change in $N_\nu$. 
In fact, this in turn results in biases of other cosmological parameters 
such as $\omega_c$ and $ \theta_s$ which are strongly correlated 
with $N_\nu$. Therefore,  $Y_p$ should be carefully treated 
in investigating cosmological constraints with future CMB data.

\begin{table}[ht]
  \begin{center}
  \begin{tabular}{l||r|r|r}
  \hline
  \hline 
  & Planck & Planck & Planck \\
  parameters & ($Y_p$: free) & ($Y_p$: BBN relation) & ($Y_p=0.24$) \\
  \hline
  $\omega_b$ & $0.02275^{+0.00025}_{-0.00028}$ & $0.02275^{+0.00026}_{-0.00027}$ & 
  $0.02273^{+0.00027}_{-0.00026}$ \\
  $\omega_c$ & $0.1108^{+0.0046}_{-0.0056}$ & $0.1101^{+0.0028}_{-0.0028}$ & 
  $0.1120^{+0.0033}_{-0.0036}$ \\
  $\theta_s$ & $1.0404^{+0.0014}_{-0.0014}$ & $1.04060^{+0.00044}_{-0.00049}$ & 
  $1.04000^{+0.00055}_{-0.00062}$ \\
  $\tau$ & $0.0881^{+0.0050}_{-0.0064}$ & $0.0881^{+0.0053}_{-0.0063}$ & 
  $0.0880^{+0.0056}_{-0.0059}$ \\
  $n_s$ & $0.964^{+0.009}_{-0.010}$ & $0.964^{+0.010}_{-0.010}$  & 
  $0.963^{+0.010}_{-0.009}$ \\
  $\ln(10^{10}A_s)$ & $3.066^{+0.016}_{-0.016}$ & $3.065^{+0.014}_{-0.015}$ & 
  $3.068^{+0.015}_{-0.015}$ \\
  $Y_p$ & $0.246^{+0.020}_{-0.018}$ & $0.2488^{+0.0027}_{-0.0027}$ & 
  --- \\
  $N_\nu$ & $3.11^{+0.33}_{-0.39}$ & $3.06^{+0.20}_{-0.19}$ &
  $3.19^{+0.24}_{-0.24}$ \\
  \hline
  $\Omega_m$ & $0.256^{+0.010}_{-0.010}$ & $0.256^{+0.009}_{-0.010}$ & 
  $0.255^{+0.009}_{-0.010}$ \\
  Age[Gyr] & $13.63^{+0.34}_{-0.31}$ & $13.67^{+0.20}_{-0.21}$ & 
  $13.56^{+0.22}_{-0.25}$\\
  $H_0$ &  $72.3^{+2.2}_{-2.4}$ & $72.0^{+1.7}_{-1.6}$ & 
  $72.7^{+1.8}_{-1.9}$ \\
  \hline
  \hline 
\end{tabular}
  \caption{Forecasts on mean values and 68\% errors of $N_\nu$ and 
  other cosmological parameters. 
  }
  \label{table:forecast}
\end{center}
\end{table}

We would like in the end to comment on how our discussion so far can be affected by
theoretical uncertainties in the recombination process  \cite{Seager:1999bc,Lewis:2006ym,Wong:2007ym,Chluba:2005uz,Chluba:2006bc,Chluba:2007yp,Switzer:2007sn,Hirata:2007sp,Switzer:2007sq}. 
Since the change of $Y_p$ can influence the recombination 
process, its uncertainties might affect the cosmological 
parameter determination in some way. Thus it may be worth 
mentioning here on the effects.
For this purpose, 
we proceed with the same analysis as have been done in Ref.~\cite{Ichikawa:2007js} 
but varying $N_\nu$ here.
Two parameters $F_H$ and $b_{He}$, 
which represent the uncertainties in the recombination modeling,
are included among other free parameters.
(See Ref.~\cite{Ichikawa:2007js} and references therein for more details). 
We impose top-hat priors, $0<F_H<2$ and $0<b_{He}<1.5$, which 
are very conservative ones, 
to take into account the uncertainties in the recombination theory.
We made the analyses for  the two cases where $Y_p$ is given from the BBN relation and $Y_p$ is treated as a free parameter.
In both cases, we found that the constraints on other cosmological 
parameters including $N_\nu$ are scarcely affected even by very 
conservative prior on $F_H$ and $b_{He}$.
The mean values are unchanged and errors increase only very 
slightly (no more than 10\% for any parameters other than $F_H$ and $b_{He}$).
Therefore we can say that the uncertainties parametrized with 
$F_H$ and $b_{He}$ do not change much our results of 
the Planck forecast discussed above.
However, we would need more understanding of uncertainties in the recombination theory  
for the precise determination of cosmological parameters 
in future CMB surveys.

\section{Summary}\label{sec:summary}

We discussed the issue of probing the effective number of neutrino
species $N_\nu$ from CMB data alone. Although a constraint on $N_\nu$
has been investigated by many authors, in most analyses, some
combinations of data set such as CMB+LSS, CMB+$H_0$, 
CMB+LSS+$H_0$ have
been used to constrain $N_\nu$.  This is partly because $N_\nu$ has
severe degeneracies in WMAP with some other cosmological parameters such as
$\omega_m$ and $h$; thus $N_\nu$ can be more constrained by
combining some data sets. However, when we combine data from LSS, some
subtleties can arise: a constraint from LSS data depends on
how we treat non-linear corrections/bias.  Furthermore, different
galaxy data lead to slightly different constraints on $N_\nu$.  In addition,
as for the Hubble prior, the prior usually adopted is $H_0 = 72 \pm 8$
which is from the result of Freedman et al \cite{Freedman:2000cf}. 
However, another
group has reported somewhat different value as $H_0 = 62.3 \pm 5.2$ 
\cite{Sandage:2006cv}.
Since different priors on the Hubble constant can give different
results, in this respect, the constraint obtained by assuming some
prior on $H_0$ should be regarded taking into account the above
uncertainty.  Taking these issues  into consideration, it may be
interesting to study a constraint on $N_\nu$ removing such subtleties,
which can be done by using CMB data alone.

In this paper, first we discussed the effects of $N_\nu$
on CMB and the issues of degeneracies with some other cosmological
parameters. Phenomenological descriptions of its effects
on the heights and the positions of acoustic peaks were also given. 
Then, in section \ref{sec:current},  
 a constraint on $N_\nu$ was studied by using CMB data
alone.  We made use of the data from WMAP5, ACBAR, BOOMERANG and CBI.  As
discussed there, although WMAP measurement is very accurate, its
precision is limited up to the 2nd peak/3rd peak. We have explicitly
shown that the information up to the 2nd/3rd peak is not enough to
constrain $N_\nu$ severely. This was demonstrated by making the analysis
with WMAP data alone, in which a sensible constraint cannot be
obtained.
However, if we include the data on small scales,
the degeneracies of $N_\nu$ with some other cosmological 
parameters can be removed to
some extent; then a stronger constraint can be obtained.  In fact, on
small scales, the amplitude is suppressed due
to the free streaming effect by increasing $N_\nu$, 
which is similar to the effects of $Y_p$
through the diffusion damping.  Thus we have studied the constraint on
$N_\nu$ assuming different priors on $Y_p$: adopting the BBN relation to derive
$Y_p$ for given $N_\nu$ and $\omega_b$, assuming $Y_p$ as a free parameter and
usual fixing of $Y_p=0.24$. Depending on the prior, the constraint
slightly changes. 
We obtained the 95 \% limits as
$0.96 \le N_\nu \le 7.94$ for the case with $Y_p$ being free, 
$1.28 \le N_\nu \le 6.84$ for $Y_p$ being fixed as $Y_p=0.24$ 
and $1.39 \le N_\nu \le 6.38$ when the BBN relation being adopted.
It should be noted that these constraints are comparable to that obtained 
using CMB+LSS in previous works.

One of the main purposes of constraining the effective number
of neutrino species using cosmological data is to check the standard
value of $N_\nu$ independently from particle physics experiments. 
Thus we primarily focus on the analysis with no prior on $N_\nu$.
However, from the viewpoint of constraining extra radiation which may be
motivated from some particle physics models, a constraint obtained 
by assuming the prior $N_\nu > 3.046$ may be interesting
 since an extra radiation always increases the value
of $N_\nu$.  In this respect, we also made an analysis adopting this prior
and obtained the constraint on the effective number of neutrino as
$N_\nu \le 6.35$ and $N_\nu \le 8.14$ 
for the cases where the BBN relation is adopted and $Y_p$ is treated as a free parameter.
 
 On the other hand, in a scenario with low-reheating
temperature, the effective number of neutrino species can be reduced.
In this case, another prior may be motivated to be assumed  for a simple
scenario of low-reheating temperature with three relativistic neutrino
species. In this regard,  we have also studied the case with the prior 
$N_\nu < 3.046$ 
and obtained the constraints as $N_\nu > 1.17$ for the cases with
$Y_p$ being assumed as a free
parameter. This can be translated into the lower bound 
on the reheating temperature as $T_{\rm reh} > 2.0$ MeV.

We have also discussed a future constraint on $N_\nu$ using the
expected data from Planck experiment.  It was shown that the
attainable constraint on $N_\nu$ from Planck is 
$2.68 \le N_\nu \le3.44$ 
at 95\% C.L.
when the BBN relation is adopted for $Y_p$, which is most stringent 
compared to the other cases.  
 Since Planck experiment can probe CMB down to smaller scales than
WMAP, Planck alone can give a stringent constraint on $N_\nu$.

The interplay between particle physics and cosmology is now becoming
more important in the era of precision cosmology.  One of such examples
is the number of neutrino species, which was investigated in this
paper.  In light of upcoming more precise observations of cosmology,
research of this kind will bring us fruitful insight for particle
physics and cosmology.

\bigskip
\bigskip

\noindent 
{\bf Acknowledgments:} 
This work is supported in part by the Japan Society for the Promotion of Science
(K.I. and T.S.), the Sumitomo Foundation (T.T.), and
the Grant-in-Aid for Scientific Research from the Ministry of
Education, Science, Sports, and Culture of Japan, No.\,18840010 (K.I.)
and No.\,19740145 (T.T.).


\end{document}